\documentclass[]{aastex631}

\shorttitle{Kinematics and chemical composition of metal-poor stars}
\shortauthors{Plotnikova et al.}

\graphicspath{{./}{figures/}}

\begin{document}

\title{Very metal-poor stars in the solar vicinity: kinematics and abundance analysis}

\author[0000-0002-7504-0950]{Anastasiia Plotnikova}
\affiliation{Dipartimento di Fisica e Astronomia, Universit\'a di Padova,I-35122, Padova, Italy}

\author[0000-0002-0155-9434]{Giovanni Carraro}
\affiliation{Dipartimento di Fisica e Astronomia, Universit\'a di Padova,I-35122, Padova, Italy}

\author[0000-0001-6205-1493]{Sandro Villanova}
\affiliation{Departamento de Astronomia, Casilla 160-C, Universidad de Concepci{\'o}n, Concepci{\'o}n. Chile}

\author[0000-0001-7939-5348]{Sergio Ortolani}
\affiliation{Dipartimento di Fisica e Astronomia, Universit\'a di Padova,I-35122, Padova, Italy}

\begin{abstract}
Very metal-poor stars contain crucial information on the Milky Way's infancy. In our previous study \citep{Plotnikova_2022} we derived a mean age of $\sim$ 13.7 Gyr for a sample of these stars in the Sun's vicinity. In this work, we investigate the chemical and kinematics properties of these stars with the goal of obtaining some insights on their origin and their parent population. 
We did not find any Al-Mg anti-correlation, which lead us to the conclusion that these stars did not form in globular clusters, while the detailed analysis of their orbital parameters reveals that these stars are most probably associated with the pristine Bulge of the Milky Way. We then sketch a scenario for the formation of the Milky Way in which the first structure to form was the Bulge through rapid collapse. The other components have grown later on, with a significant contribution of accreted structures. 
\end{abstract}


\section{Introduction} \label{sec:intro}
Very metal-poor stars are one of the best candidates to study the formation and evolution of the Universe and its Galactic components. The study of their chemistry and kinematics can help us to understand the chemical and dynamical properties of the first Population III supernovae. The origin of these stars contains important information about the Milky Way formation history.

Theoretical simulations of the Galaxy formation (\citet{Bullock2005}) have shown that the Halo bears the signatures of the Milky Way's assembly from smaller “building block” galaxies. Recent astrometric studies have shown the existence of kinematic signatures that indicate past accretion events (\citet{Belokurov2018}; \citet{Myeong2019}; \citet{Yuan2020}). That means that the kinematics of very metal-poor stars is an important testbed of the Galaxy formation and evolution theories. In fact, stars with very low metal abundances are potential members of accreted dwarf galaxies and/or clusters.

Two main theories were suggested for the Milky Way formation. The first theory is the hierarchical scenario which tells us that our Galaxy was formed through the hierarchical merging of smaller dark halos. The accretion of baryonic matter occured later. The Bulge was formed first, followed by the Thin Disk. The Thick Disk could have been produced by the kinetic heating induced by small/medium mass engulfing companions.

The presence of stellar streams in the stellar Halo supports the hierarchical scenario (\citet{Helmi_2002}, \citet{Ibata_2002}): they are tidal remains of past merging events (\citet{Grillmair_2017}). Nowadays, we have a number of confirmed accreted events such as Sagittarius dwarf galaxy, Gaia-Enceladus-Sausage, Sequoia, Helmi stream, Thamnos, while other streams were found to be associated with accreted globular clusters (\citet{Koppelman_2019}). The Halo of the Galaxy could have mostly be assembled in a sequence of minor mergers. Furthermore, the fact that the merging frequency is increasing with redshift, tells us that in the infant Galaxy merging events were quite common. That lends further support to the hierarchical scenario.

The second and most successful scenario is Secular Evolution (SE), with slow but continuous external matter accretion. According to this theory, the Bulge of the Galaxy was formed due to the accretion of disk matter through bar instabilities. This scenario is in agreement with the observed color gradient and the relation between the color of the bulge and disk studied through statistical analysis of 257 spiral galaxies (\citet{Gadotti_2001}). SE is also supported by the relation between the bulge and disk masses and radii (\citet{Courteau_1996}).

Metal-poor stars, being mostly ancient, are ideal probes of the early evolutionary phases of the Universe. Because of their low metallicity, they are linked to the most pristine star formation episodes in the Universe.

In this work, we build on \citep{Plotnikova_2022} and study the kinematics and abundances of a sample of very metal-poor stars in the solar vicinity.
Therefore, the layout of this study is as follows.
In Section~2 we discuss the data upon which our study is based.
Section~3 and 4 are focused on the chemistry and kinematics of our stars, respectively. The results of our investigation are detailed in Section~5, while Section~6, finally, summarizes the outcome of this study and highlights future perspectives.

\section{Data}
As the main target for this investigation, we chose the data set of 28 very metal-poor stars for which we derived precise ages (\cite{Plotnikova_2022}). These stars are of particular interest because they are very metal-poor and their average age is $13.7 \pm 0.4$ Gyr. As a result, the stars in this sample are valuable candidates to study the early stages of the Galaxy 
formation.

\subsection{Distance determination}\label{Distance determination}
Distance is one of the most important parameters in dynamics studies. In \cite{Plotnikova_2022} we compared the best distances, estimated out of four different techniques: Gaia DR3 parallaxes (\cite{GaiaDR3_2022}), Gaia EDR3 (\cite{GaiaEDR32021}) corrected by \cite{Lindegren2021}, the distances derived by \cite{Bailer-Jones2021}, and \cite{Queiroz_2019} (\texttt{StarHorse}). Our choice was made based on the best fit of the data with the isochrones in the color-magnitude diagram. As a result, we found that the best distances are those obtained directly from Gaia DR3 parallaxes (\cite{Plotnikova_2022}).

\subsection{Astrometric parameters}
With the goal of calculating orbits and orbital parameters, we extracted coordinates, radial velocities, and proper motion components from the Gaia DR3 archive (\cite{GaiaDR3_2022}). Uncertainties for velocity components are shown in Tab.\ref{tab:Unc GEDR3 radial velocity}, \ref{tab:Unc GEDR3 astrometry}. Position and velocity of the stars in the Equatorial Coordinates were transformed to Galactocentric coordinates by \texttt{astropy}\footnote{\url{https://docs.astropy.org/en/stable/_modules/astropy/coordinates/builtin_frames/galactocentric.html astropy}} python package.

\begin{table}[]
    \centering
    \begin{tabular}{lccc}
    \hline
    \hline
        Data product or source type&\multicolumn{3}{c}{Typical uncertainty, km s$^{-1}$}\\
        &G$_{RVS} <$ 8&G$_{RVS}$ = 10&G$_{RVS}$ = 11.75\\
        \hline
        Median radial velocity over 22 months &0.3&0.6&1.8\\
        Systematic radial velocity errors&$<$ 0.1&-&0.5\\
        \hline
    \end{tabular}
    \caption{Uncertainties of Gaia Early Data Release 3 proper motions}
    \label{tab:Unc GEDR3 radial velocity}
\end{table}

\begin{table}[]
    \centering
    \begin{tabular}{lcccc}
        \hline
        \hline
        Data product or source type&\multicolumn{4}{c}{Typical uncertainty}\\
        &G $<$ 15&G = 17 &G = 20 &G = 21 \\
        \hline
        Five-parameter astrometry&&&&\\
        \hline
        position, mas&0.01 - 0.02&0.05&0.4&1\\
        parallax, mas&0.02 -
        0.03&0.07&0.5&1.3\\
        proper motion, mas yr$^{-1}$&0.02 - 0.03&0.07&0.5&1.4\\
        \hline
        Six-parameter astrometry&&&&\\
        \hline
        position, mas&0.02 - 0.03&0.08&0.4&1\\
        parallax, mas&0.02 -     0.04&0.1&0.5&1.4\\
        proper motion, mas yr$^{-1}$ & 0.02 - 0.04&0.1&0.6&1.5\\
        \hline
    \end{tabular}
    \caption{Uncertainties of Gaia Early Data Release 3 astrometry \citet{GaiaEDR32021}}
    \label{tab:Unc GEDR3 astrometry}
\end{table}

\section{Chemistry}
All our stars are from Hamburg vs. ESO R-process Enhanced Star (HERES) survey. The chemical abundances were spectroscopically studied by \cite{Barklem2005}. The spectrum synthesis assumes LTE and a 1D plane-parallel model of the atmosphere, where turbulence is modeled through the classical micro-turbulence and macro-turbulence parameters. Their snapshot spectra cover a wavelength range of 3760 – 4980 Å and have an average signal-to-noise ratio of S/N $\sim$ 54 per pixel over the entire spectral range. A $2''$ slit is employed giving a minimum resolving power of $R \approx$ 20000. From the "snapshot" spectra the elemental abundances of moderate precision (absolute r.m.s. errors of order 0.25 dex, relative r.m.s. errors of order 0.15 dex) have been obtained for 22 elements: C, Mg, Al, Ca, Sc, Ti, V, Cr, Mn, Fe, Co, Ni, Zn, Sr, Y, Zr, Ba, La, Ce, Nd, Sm, and Eu. The abundances used in this work are listed in Tab.\ref{tab:Chemistry}.

\begin{table}[]
    \centering
    \begin{tabular}{lcccccccccr}
    \hline
    \hline
        ID & [Mg/Fe] & [Al/Fe]$_{LTE}$ & [Al/Fe]$_{NLTE}$ & [Ca/Fe] & [Cr/Fe] & [Ti/Fe] & [Ni/Fe] & \textbf{[Fe/H]} & \textbf{age} & \textbf{Origin*}\\
        & dex & dex & dex & dex & dex & dex & dex & dex & Gyr &\\
        \hline
        HE\_0023-4825 & 0.22 & -1.01 & -0.51 & 0.26 & 0.00 & 0.31 & -0.01 & -2.06 & 14.62 $\pm$ 0.11 & Pr.B.\\
        HE\_0109-3711 & - & - & - & 0.42 & -0.06 & 0.38 & - & -1.91 & 11.78 $\pm$ 0.04 & 5G/GE\\
        HE\_0231-4016 & 0.22 & -1.09 & -0.50 & 0.36 & -0.11 & 0.25 & -0.14 & -2.08 & 14.37 $\pm$ 0.47 & Pr.B.\\
        HE\_0340-3430 & 0.19 & -1.00 & -0.61 & 0.36 & -0.16 & 0.25 & -0.26 & -1.95 & 13.32 $\pm$ 0.13 & 5G/GE\\
        HE\_0430-4404 & 0.29 & -0.97 & -0.57 & 0.33 & -0.02 & 0.32 & 0.01 & -2.07 & 10.15 $\pm$ 1.38 & Th.1\\
        HE\_0447-4858 & 0.24 & -0.81 & -0.42 & 0.24 & 0.09 & 0.28 & 0.33 & -1.69 & 13.28 $\pm$ 0.20 & 5G/GE\\
        HE\_0501-5139 & 0.19 & - & - & 0.34 & -0.24 & 0.39 & -0.13 & -2.38 & 8.93 $\pm$ 0.03 & Th.1\\
        HE\_0519-5525 & 0.41 & -0.76 & -0.26 & 0.37 & -0.19 & 0.37 & 0.13 & -2.52 & 13.46 $\pm$ 0.06 & 5G/GE\\
        HE\_0534-4615 & 0.22 & -0.87 & -0.37 & 0.28 & -0.18 & 0.19 & -0.20 & -2.01 & 14.91 $\pm$ 0.49 & Pr.B.\\
        HE\_1052-2548 & 0.16 & -0.48 & -0.11 & 0.27 & -0.12 & 0.45 & 0.03 & -2.29 & 15.70 $\pm$ 0.14 & Pr.B.\\
        HE\_1105+0027 & 0.47 & -0.89 & -0.33 & 0.47 & 0.05 & 0.32 & -0.29 & -2.42 & 12.06 $\pm$ 0.35 & Th.1\\
        HE\_1225-0515 & 0.18 & -1.00 & -0.62 & 0.27 & -0.08 & 0.34 & -0.16 & -1.96 & 14.64 $\pm$ 0.03 & Pr.B.\\
        HE\_1330-0354 & 0.32 & -0.93 & -0.46 & 0.40 & -0.05 & 0.54 & -0.08 & -2.29 & 13.49 $\pm$ 0.34 & SimD\\
        HE\_2250-2132 & 0.31 & -1.07 & -0.57 & 0.29 & -0.12 & 0.35 & 0.05 & -2.22 & 13.19 $\pm$ 0.11 & SimD\\
        HE\_2347-1254 & 0.29 & -0.74 & -0.42 & 0.37 & -0.13 & 0.39 & -0.16 & -1.83 & 14.56 $\pm$ 0.03 & Pr.B.\\
        HE\_2347-1448 & 0.13 & -1.11 & -0.64 & 0.21 & -0.05 & 0.31 & 0.27 & -2.31 & 8.63 $\pm$ 0.07 & 5G/GE\\
        HE\_0244-4111 & 0.34 & - & - & 0.37 & -0.19 & 0.30 & 0.00 & -2.56 & 12.75 $\pm$ 0.51 & Th.1\\
        HE\_0441-4343 & 0.32 & -1.09 & -0.59 & 0.18 & -0.36 & 0.27 & 0.07 & -2.52 & 10.41 $\pm$ 0.50 & Th.1\\
        HE\_0513-4557 & 0.34 & - & - & 0.21 & - & - & - & -2.79 & 12.42 $\pm$ 0.50 & OrNC\\
        HE\_0926-0508 & 0.28 & -0.90 & -0.49 & 0.37 & -0.14 & 0.49 & -0.06 & -2.78 & 14.79 $\pm$ 0.51 & Pr.B.\\
        HE\_1006-2218 & - & -0.79 & -0.46 & 0.32 & -0.11 & 0.48 & - & -2.69 & 12.90 $\pm$ 0.51 & Th.2\\
        HE\_1015-0027 & 0.35 & -1.00 & -0.67 & 0.41 & -0.24 & 0.55 & -0.10 & -2.66 & 15.20 $\pm$ 0.55 & Pr.B.\\
        HE\_1126-1735 & 0.31 & -1.06 & -0.46 & 0.32 & -0.29 & 0.33 & -0.04 & -2.69 & 9.44 $\pm$ 0.50 & GE\\
        HE\_1413-1954 & - & - & - & 0.33 & - & - & - & -3.22 & 14.20 $\pm$ 0.82 & Pr.B.\\
        HE\_2222-4156 & 0.42 & -0.93 & -0.42 & 0.35 & -0.21 & 0.33 & -0.02 & -2.73 & 13.50 $\pm$ 0.52 & Th.1\\
        HE\_2325-0755 & 0.31 & -1.02 & -0.42 & 0.46 & -0.29 & 0.35 & -0.18 & -2.85 & 13.50 $\pm$ 0.51 & SimD\\
        \hline
    \end{tabular}
    \begin{footnotesize}
    *GE - Gaia-Enceladus/Sausage, Th.1 - Thamnos 1, Th.2 - Thamnos 2, Pr.B. - primordial bulge, 5G/GE - 5 stars group, high probability to belong to the GE, SimD - similar to the disk kinematics, OrNC - origin is not clear, stars have halo kinematics and chemistry.
    \end{footnotesize}
    \caption{Chemistry}
    \label{tab:Chemistry}
\end{table}

\subsection{Non-LTE corrections}
As mentioned above, spectral analysis in \cite{Barklem2005} was done assuming only the LTE model. However, aluminum abundance was obtained from the lines in the ultra-violet part of the spectrum that are significantly affected by Non-LTE effects (\cite{Nordlander_2017}). To improve the precision of the chemical parameters we applied Non-LTE correction from \cite{Nordlander_2017}. The correction was implemented for each star according to its effective temperature ($T_{eff}$), metallicity ($[Fe/H]$) and gravity ($log g$). The resulting obtained corrections are on average around 0.5 dex (Tab.\ref{tab:Chemistry}).

We also checked Non-LTE corrections for all other elements by means of MPIA webtool database\footnote{\url{https://nlte.mpia.de}} but in all cases, corrections are less than 0.1 dex and were considered negligible.


\section{Kinematics}

We obtained orbits and orbital parameters for all stars in the data set using parallaxes, proper motion and radial velocities from Gaia DR3 by numerical calculation and adopting a model of Galactic potentials. The results are then analyzed to obtain insights on the stars' origin.

\subsection{Galactic axisymmetric potential}\label{Gal pot}
We corrected the velocities of the stars by the velocity of the Sun with respect to the Galactic center which is computed independently from the velocity of the local standard of rest (\cite{Reid_2004_vSun}, \cite{GRAVITY_Collaboration_2018}, \cite{Drimmel_2018_vSun}):

\begin{equation}
    v_{R\odot} = -12.9 \pm 3.0 \text{ km s}^{-1}
\end{equation}
\begin{equation}
    v_{\phi\odot} = 245.6 \pm 1.4 \text{ km s}^{-1}
\end{equation}
\begin{equation}
    v_{Z\odot} = 7.78 \pm 0.09 \text{ km s}^{-1}
\end{equation}
where the distance between the Galactic center and the Sun is $R_0 = 8.122 \pm 0.033$ kpc (\cite{GRAVITY_Collaboration_2018}).
 
We used the \cite{McMillan_2017} Galactic potential without bar implementation in \texttt{galpy}\footnote{\url{http://github.com/jobovy/galpy}} (\cite{Bovy_2015_galpy}) to calculate orbits and orbital parameters such as the total energy ($E_{n}$), eccentricity ($e$), apo-center ($R_{apo}$), and peri-center ($R_{peri}$).

\subsection{Galactic non-axisymmetric potential}\label{NAP}
To take into account the effect of the bar we replaced the axisymmetric bulge from \cite{McMillan_2017} with a non-axisymmetric elongated bar/bulge component. As a model for the  bar/bulge structure, we used the rotating ellipsoid (\cite{Chemel_2018}, \cite{Yeh_2020}) with density distribution given by Ferrers formula:(\cite{Bovy_2015_galpy}): 

\begin{equation}
    \rho(x,y,z)=
    \left\{  
       \begin{array}{lcl}  
        \rho_{c}(1-m^{2})^{2} & , & \text{if } m<1, \\  
        0 & , & \text{if } m>1, \\  
       \end{array}   
    \right.  
\end{equation}

where $m^{2}=\frac{x^2}{a^2}+\frac{y^2}{b^2}+\frac{z^2}{c^2}$. For the parameters, we used $a=5$ kpc, $b=2$ kpc, $c=1$ kpc which gives a good approximation for the observed bar/bulge component in the center of the Milky Way (\cite{Portail_2015}, \cite{Portail_2017}). The central density of the bar is given by $\rho=\frac{105}{32\pi}\frac{GM_b}{abc}$, where the mass of the bar $M_b=1.88 \times 10^{10}$ M$_{\odot}$ (\cite{Portail_2017}).

The Milky Way bar is rotating around the Galactic center with constant angular velocity $\Omega_b$. Also, its major axis is tilted with respect to the Sun.

Recent measurements of bar angular velocity give the following results: $\Omega_b=41\pm3$ km s$^{-1}$ kpc$^{-1}$ (\cite{Sanders_2019}), $\Omega_b=37.5$ km s$^{-1}$ kpc$^{-1}$ (\cite{Clarke_2019}), $\Omega_b=40$ km s$^{-1}$ kpc$^{-1}$ (\cite{Sormani_2015}). In this work we used bar angular velocity $\Omega_b=40$ km s$^{-1}$ kpc$^{-1}$ which is in a good agreement with \cite{Portail_2015}, \cite{Portail_2017}, and \cite{Bovy_2019}. For the bar orientation relative to the Sun we used $\phi=28^o$ which is in good agreement with recent estimates (\cite{Wegg_2013}, \cite{Bland-Hawthorn_2016}, and 
 \cite{Portail_2017}). 

\subsection{Uncertainty}
To estimate uncertainties for each derived orbital parameter we perform a Monte Carlo analysis using 1000 random draws on the input parameters: parallax, radial velocity and proper motion. We considered errors in right ascension (RA) and declination (DEC) to be negligible compared to other parameters. As for proper motion, we used the multivariate normal distribution which takes into account the correlation between both components\footnote{\url{https://gea.esac.esa.int/archive/documentation/GEDR3/Gaia_archive/chap_datamodel/sec_dm_main_tables/ssec_dm_gaia_source.html}}. For parallax and radial velocity we used just normal distribution.

Since we use parallaxes, the resulting distribution of each orbital parameter is not Gaussian. Therefore, we derived for each orbital parameter a median value and a confidence interval of 68\%.

\section{Results}\label{results}
We studied the chemical composition and kinematics of our data set of 28 metal-poor stars and the correlation with their age.

\subsection{Chemistry}
\begin{figure}
\centering
\includegraphics[scale=0.5]{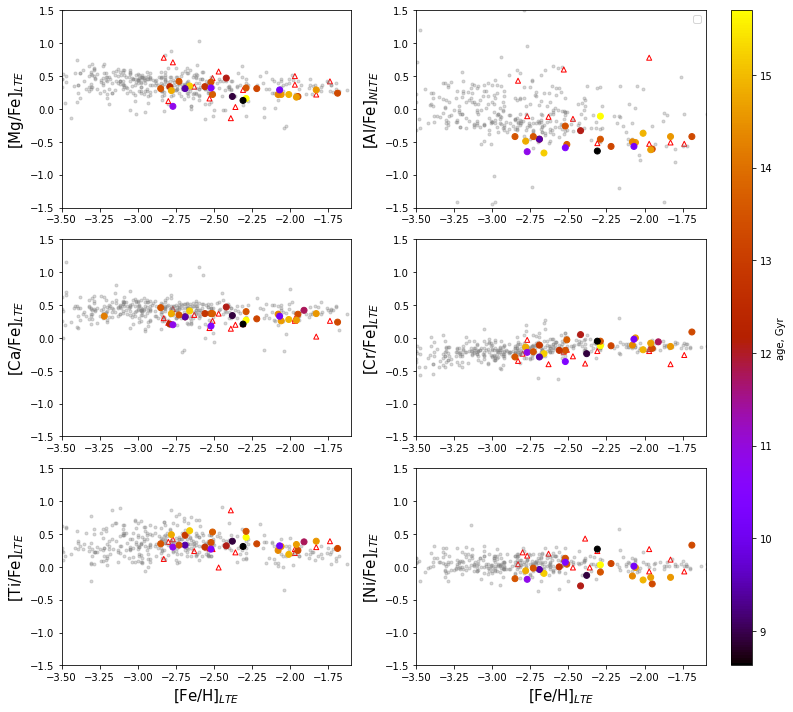}
\caption{Correlation between iron and other chemical elements color-coded with the age. For comparison we show Bulge (red open triangles, \cite{Howes_2016}) and Halo (grey circles, \cite{Yong_2013}, \cite{Roederer_2014}) stars of the Milky Way. For all sources Al correction from \cite{Nordlander_2017} was applied.}
\label{fig:Chemistry c_age}
\end{figure}

In Fig.\ref{fig:Chemistry c_age} we compared the chemical composition of our data set with bulge (red open triangles, \cite{Howes_2016}) and halo (grey circles, \cite{Yong_2013}, \cite{Roederer_2014}) stars of the Milky Way. Inspecting this figure one can readily see that all our stars are in good agreement with the typical trends for  Milky Way stars. This holds except for aluminum, whose position is shifted on average by 0.2 dex. For all stars we applied non-LTE correction from \cite{Nordlander_2017} that play important role in aluminum abundance.


\begin{figure}
\centering
\includegraphics[scale=0.5]{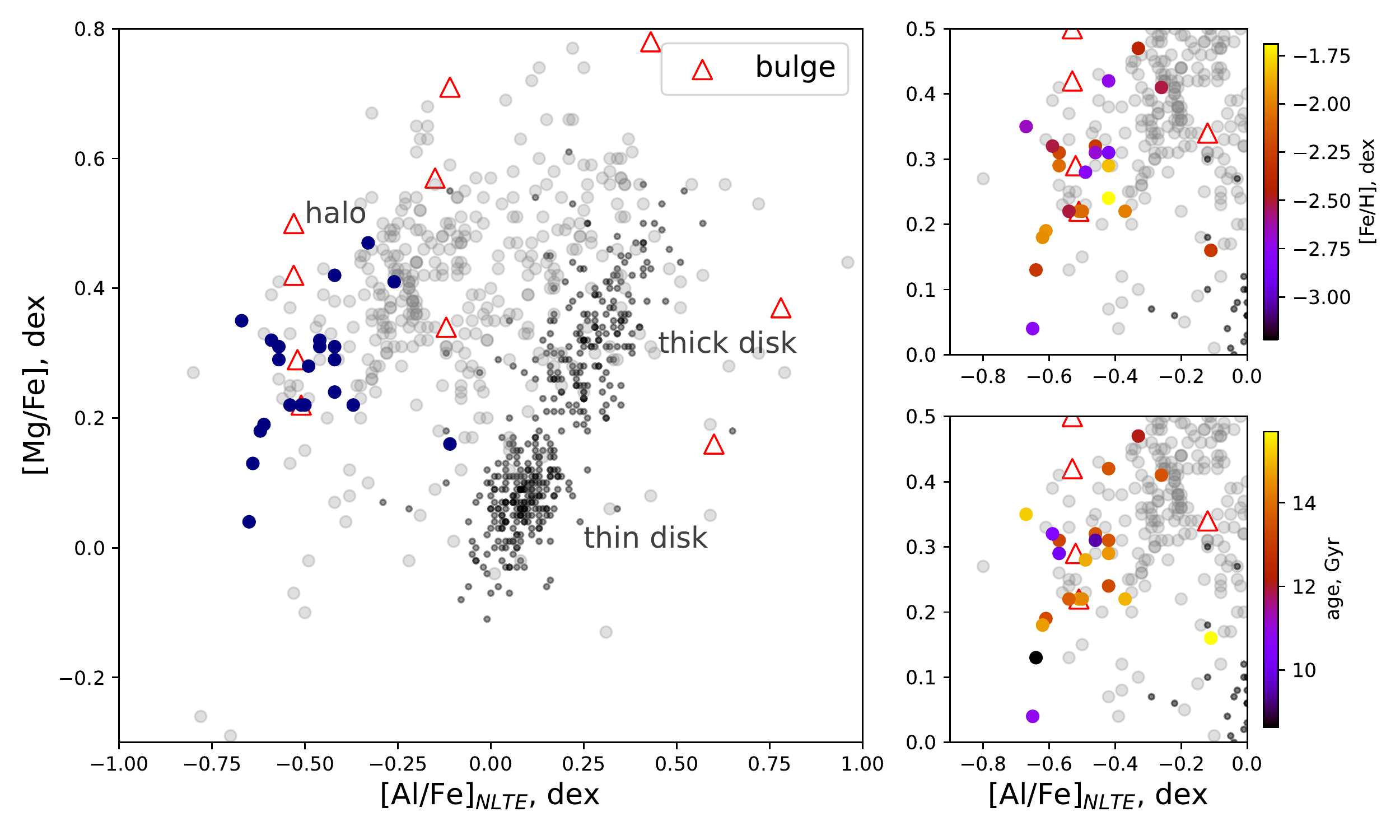}
\caption{Al-Mg correlation. \textit{Top right}: color-coded with metallicity, \textit{bottom right}: color coded with age. The background: disk stars (\textit{black dotes}, \cite{Fulbright_2000}, \cite{Reddy_2003}, \cite{Simmerer_2004}, \cite{Reddy_2006}, \cite{Francois_2007}, \cite{Johnson_2012}, \cite{Johnson_2014}), halo stars (\textit{grey points}, \cite{Yong_2013}), \cite{Roederer_2014}, bulge stars (\textit{red open triangles}, \cite{Howes_2016}).}
\label{fig:AlMg}
\end{figure}

To separate stars from the Halo, the Thin, and the Thick disk stars we used the Al versus Mg map (Fig.\ref{fig:AlMg}). For comparison we used for Disk: \cite{Fulbright_2000}, \cite{Reddy_2003}, \cite{Simmerer_2004}, \cite{Reddy_2006}, \cite{Francois_2007}, \cite{Johnson_2012}, \cite{Johnson_2014}; for Halo: \cite{Yong_2013}, \cite{Roederer_2014}; and for Bulge: \cite{Howes_2016}. All our stars having measures for both elements occupy the expected region for halo stars. At odds with globular clusters' stars, our sample stars do not show any Al-Mg anti-correlation (\cite{Lucey_2022}, \cite{Sestito_2022}). 

We can therefore conclude that our stars possess the typical chemical composition of the Milky Way Halo field stars. The lack of any anti-correlation seems to exclude an origin inside globular clusters.

\subsection{Kinematics}
\begin{figure}
\centering
\includegraphics[scale=0.5]{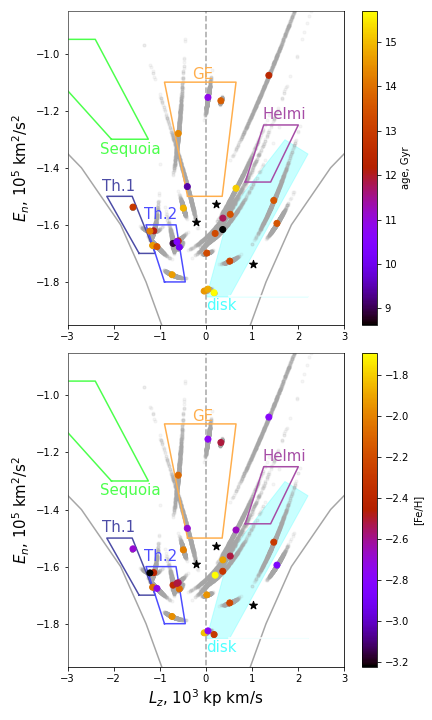}
\caption{$E_n - L_z$ diagram with the loci of the biggest known accretion events: Gaia-Enceladus/Sausage (GE) (\textit{orange}), Helmi stream (\textit{purple}), Sequoia (\textit{green}), Thamnos 1, 2 (\textit{darkblue, blue}); and disk (\textit{cyan}). Black star symbols indicate Vandenberg halo stars \cite{VandenBerg2014}. The colored points are mean values of stars under investigations, while grey trails indicate their uncertainties.}
\label{fig:ELz Fe age}
\end{figure}

\begin{figure}
\centering
\includegraphics[scale=0.5]{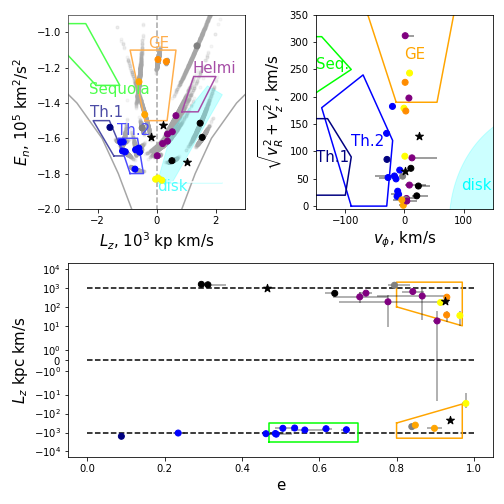}
\caption{Points are color-coded with the position in $E_n - L_z$ diagram: Gaia-Enceladus/Sausage (GE) (\textit{orange}), Sequoia (\textit{green}), Thamnos 1, 2 (Th.) (\textit{dark blue, blue}), Sagittarius (\textit{red}), disk (\textit{black}), group of 5 stars with similar parameters (\textit{purple}), low energy stars (\textit{yellow}), Vandenberg stars \cite{VandenBerg2014} (\textit{black stars}). }
\label{fig:eLz}
\end{figure}

To get additional insight on the origin of these stars we explore their kinematical properties. To this aim, in Fig.\ref{fig:ELz Fe age} we compared our results with \cite{Koppelman_2019}. We used the same \cite{McMillan_2017} potential as in \cite{Koppelman_2019} but, after checking that there are no differences,  we adopted the most recent estimates of the Sun's velocity and distance to the Galactic center (Sec.\ref{Gal pot}). Fig.\ref{fig:ELz Fe age} shows that some of the stars share the same kinematics as well-known accretion events: the Gaia-Enceladus/Sausage system (\textit{orange}; \cite{Belokurov2018}, \cite{Helmi_2018}) and Thamnos 1,2 (\textit{darkblue},\textit{blue};  \cite{Koppelman_2019}). This result is also supported by the velocity and $L_{z}$-eccentricity diagrams (Fig.\ref{fig:eLz}).

Besides, we can notice a group of 5 stars below the Gaia-Enceladus/Sausage locus with similar orbital parameters, age, and metallicity, and location inside the sphere with radius R$\sim$1 kpc. In some sources, Gaia-Enceladus/Sausage locus is extended to  lower energies (\cite{Koppelman_2019}) which gives a higher probability for this 5 stars group to be indeed part of Gaia-Enceladus/Sausage. Moreover, there are three stars (Fig.\ref{fig:eLz}, \textit{black dotes}) that exhibit disk kinematics, although their chemistry looks more similar to the halo .

Our stars exhibit both prograde and retrograde motion and they cover almost the entire eccentricity range (Fig.\ref{fig:eLz}, \textit{lower panel}). Additionally, all stars are distributed around $v_{\phi}=0$ km/s which indicates a typical Halo  kinematics (Fig.\ref{fig:eLz}, \textit{upper right panel}). They might be part either of Gaia-Enceladus/Sausage (GE) (\textit{orange}), or of Thamnos 1, 2 (\textit{dark blue, blue}) accretion events. All diagrams in Fig.\ref{fig:eLz} are color-coded according to their position in the $L_{z}-E_{n}$ diagram (Fig.\ref{fig:eLz}, \textit{upper left panel}) 

\begin{figure}
\centering
\includegraphics[scale=0.5]{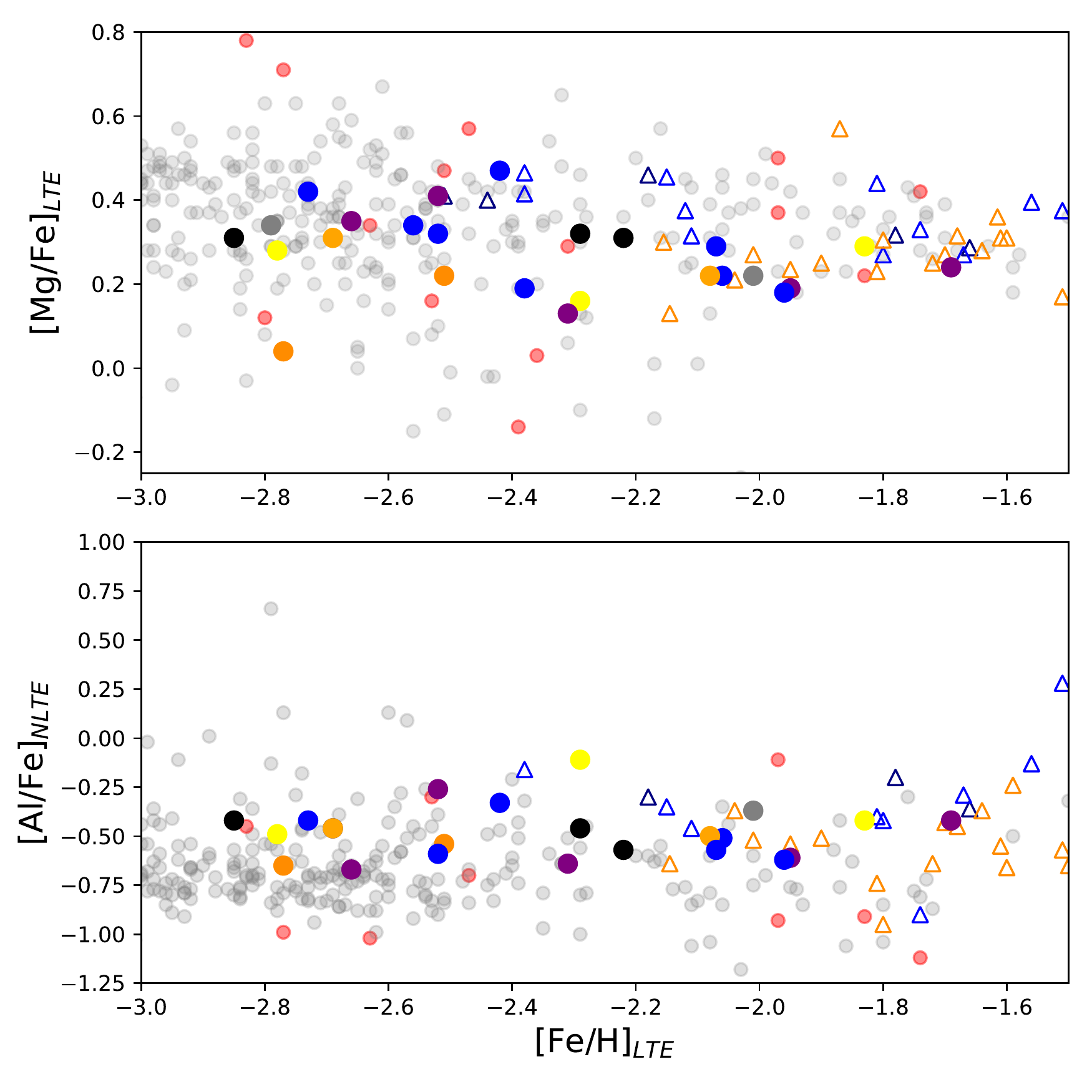}
\caption{Correlation between Mg, Al and Fe. For comparison, we showed bulge (red circles, \cite{Howes_2016}) and halo (grey circles, \cite{Yong_2013}, \cite{Roederer_2014}) stars of the Milky Way. Points are color-coded with the position in $E_n - L_z$ diagram: Gaia-Enceladus/Sausage (GE) (\textit{orange}), Sequoia (\textit{green}), Thamnos 1, 2 (Th.) (\textit{dark blue, blue}), Sagittarius (\textit{red}), disk (\textit{black}), group of 5 stars with similar parameters (\textit{purple}), low energy stars (\textit{yellow}). \textit{Circles} - stars from this work, \textit{open triangles} - stars from known accretion events \cite{Koppelman_2019}).}
\label{fig:MgAlFe}
\end{figure}

However, the kinematic analysis alone is not sufficient to identify the origin of these stars and decipher whether they are associated or not with a specific accreted structure. To check the possible membership of \textit{blue} and \textit{orange} stars to Gaia-Enceladus/Sausage (GE) (\textit{orange}) or Thamnos 1, 2 (\textit{dark blue, blue}) we need to combine kinematics with chemistry. In Fig.\ref{fig:MgAlFe} we can see that both Milky Way stars and accretion events possess similar chemistry. 

\begin{figure}
\centering
\includegraphics[scale=0.5]{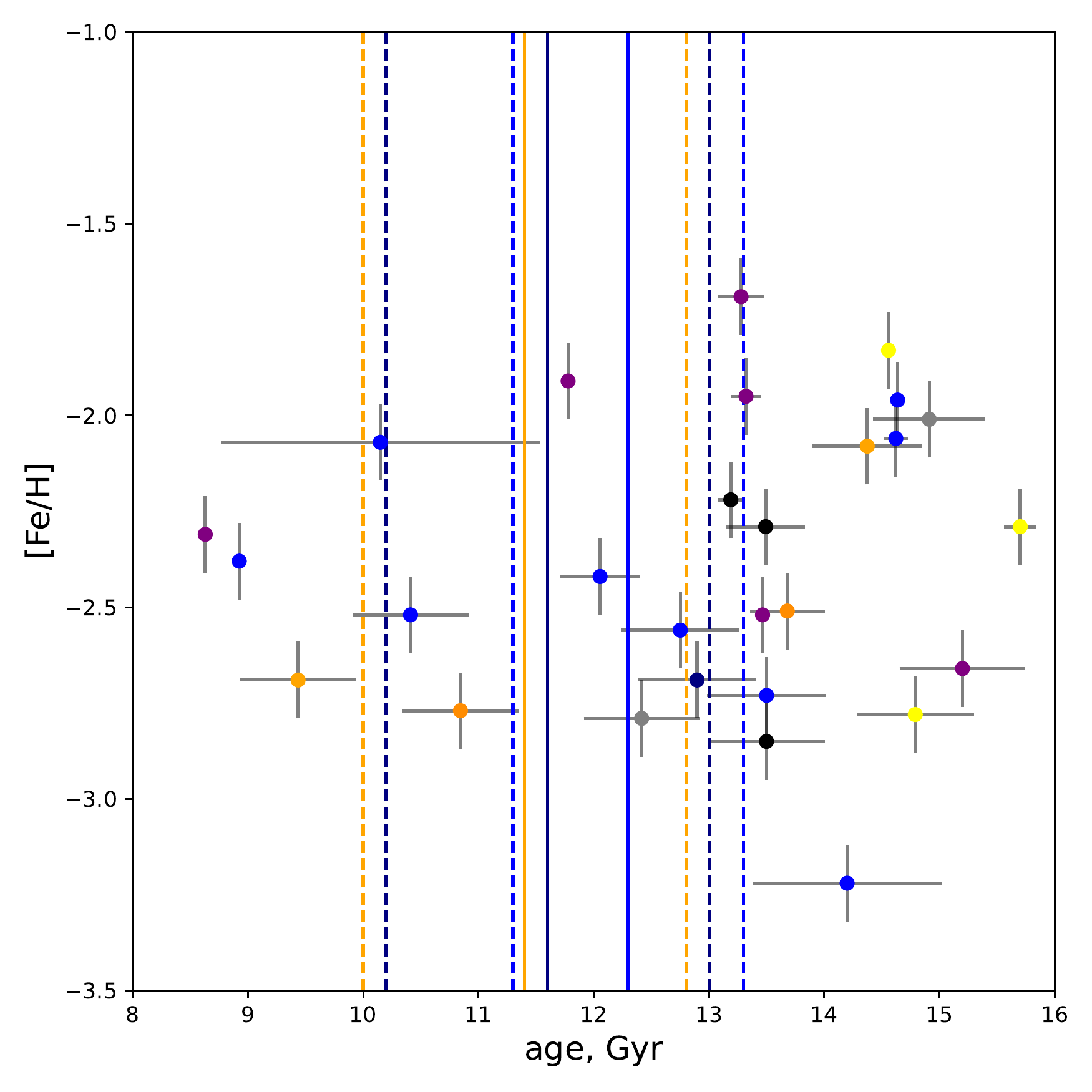}
\caption{Age-metallicity relation with the time bars of the Gaia-Enceladus/Sausage (GE) (\textit{orange}), Thamnos 1, 2 (\textit{dark blue, blue}) formation (\cite{Ruiz-Lara_2022}). Solid lines are mean values and dashed lines represent uncertainties.}
\label{fig:age of accretion events}
\end{figure}

Overall, our metal-poor stars have chemistry compatible with accretion event trends. Additionally, we compared the age of the stars with the age of each accretion event (Fig.\ref{fig:age of accretion events}) and a part of the star's ages fall into the range of the beginning of the accreting blocks' formation. But some of the stars from this work are older and can be excluded from being accreted through known accretion events.

\subsection{Orbital parameters}
\begin{figure}
\centering
\includegraphics[scale=0.5]{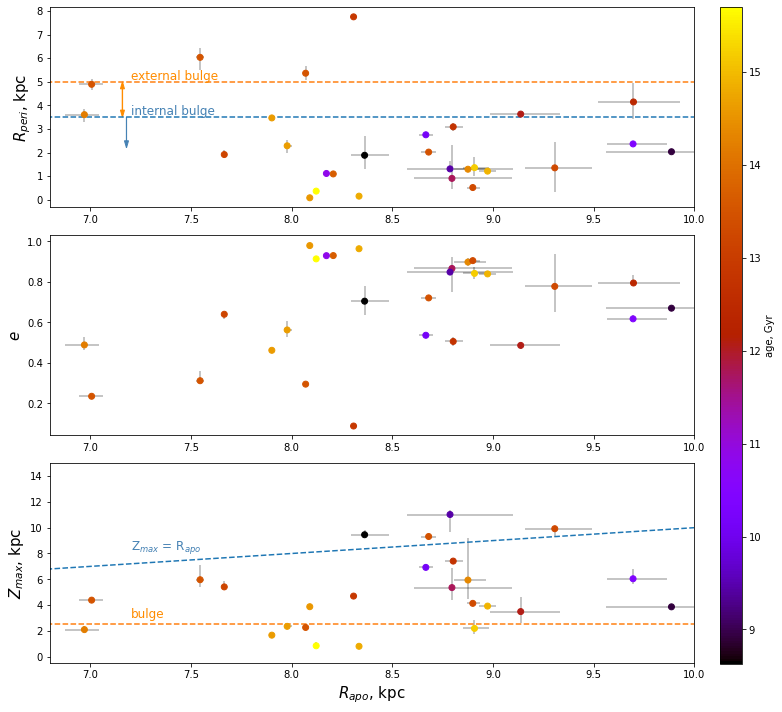}
\caption{Correlation between eccentricity ($e$), pericentre ($R_{peri}$), maximum height under the galactic plane ($Z_{max}$) and apocentre ($R_{apo}$) color-coded with age.}
\label{fig:orb param age}
\end{figure}

In Fig.\ref{fig:orb param age} we presented the correlation between eccentricity ($e$), peri-centre ($R_{peri}$), maximum height above the galactic plane ($Z_{max}$) and apo-centre ($R_{apo}$) color-coded with age. From the upper plot in Fig.\ref{fig:orb param age} we can notice that almost all stars in our data set cross the bulge region. That means that these stars' kinematics is perturbed by the bar/bulge potential. One can also speculate that they can be a part of the pristine bulge of the Milky Way. From the middle plot in Fig.\ref{fig:orb param age} we can see that the most distant stars exhibit on average more elongated orbits. The lower plot in Fig.\ref{fig:orb param age} shows that most of the stars have $R_{apo}>Z_{max}$, i.e. they lie close to the Galactic plane. Another intriguing evidence in this plot is that one can readily see a clear trend in age: the oldest stars (yellow) are located lower to the plane while the younger (redder) stars lie closer to the dashed line $R_{apo}=Z_{max}$.

\begin{figure}
\centering
\includegraphics[scale=0.5]{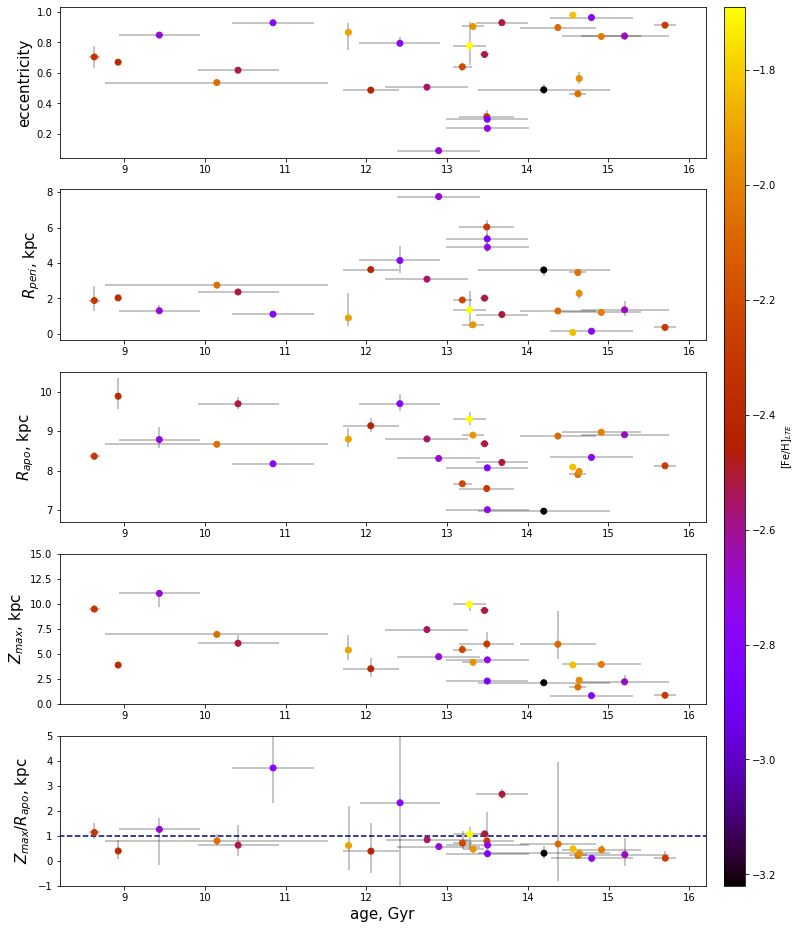}
\caption{Correlation between eccentricity ($e$), pericentre ($R_{peri}$), apocentre ($R_{apo}$), maximum height under the galactic plane ($Z_{max}$), $Z_{max}/R_{apo}$ and age color-coded with metallicity.}
\label{fig:orb param Fe}
\end{figure}

By inspecting Fig.\ref{fig:orb param Fe} we can highlight also the following trends:
\begin{itemize}
    \item [1.] Eccentricity is increasing with increasing age.
    \item [2.] The older stars orbit closer to the Galactic center and the Galactic disk.
\end{itemize}

\subsubsection{Very old stars}
Let us draw our attention to stars with ages larger than $13.5$ Gyr. Their age implies they have a very low probability to be part of any well-known accretion events (Fig.\ref{fig:age of accretion events}). Additionally, their age is very close to the age of the Universe and the very first instants of the Milky Way formation. We then speculate that they can have formed inside the very first "building blocks" of the Milky Way and therefore were part of the infant bulge at a time when the Disk had yet to form (\cite{Carraro_1999}). 

Moreover, all of them have orbits close to the Galactic Disk and pass close to the Galactic center while crossing the Galactic bulge. This might lend further support to a scenario in which these stars were formed in a pristine Bulge and later displaced in larger eccentricity orbits by some migration phenomenon. The same result was obtained by \cite{Grenon_1985} for metal-rich bulge stars in the solar vicinity. However, closeness to the galactic plane is rather the result of the observational limitations (all of our stars lie inside 5 kpc sphere around the Sun) than an intrinsic feature. We consider that with better observations there is a probability to find old metal-poor stars passing through the bulge in elongated orbits in all directions from the Galactic center.

\begin{figure}
\centering
\includegraphics[scale=0.5]{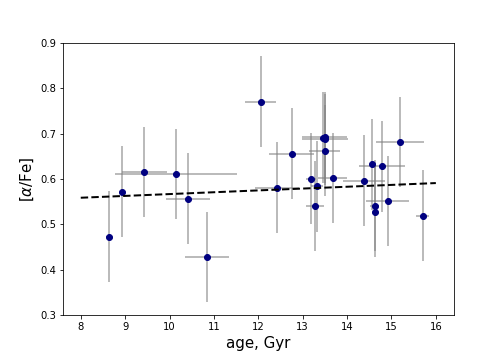}
\caption{Correlation between $\alpha$-elements and age. \textbf{The dashed line shows the weighted fit}.}
\label{fig:alpha age}
\end{figure}

Moving back to chemistry, stars under investigation are $\alpha$-enhanced, and close to typical of today's Milky Way Halo stars. Speaking about bulge stars, according to \cite{Howes_2016} data, chemical composition of Bulge stars is more disperse than Halo stars but distributed around the same mean that makes it impossible to distinguish between Bulge and Halo stars. Still, and interestingly, for all stars in our data set there is a clear correlation between the abundance of $\alpha$-elements and age (Fig.\ref{fig:alpha age}) with weighted Person correlation coefficient equal to 0.24. This is due to the fact that for low metallicity stars the main source of $\alpha$-elements is supernovae of type II (SNe II). An increasing $\alpha$-abundance can be a signature of an environment where gas density and star formation were quite high. This seems to be another argument in favour of the idea  that our oldest stars were associated with the ancient bulge. 

\subsection{Non-axisymmetric bar/bulge potential}
Since most of our stars orbit close to the Galactic center ($R_{peri}<3.5$ kpc) we additionally studied their kinematics in a non-axisymmetric bar/bulge potential (Sec.\ref{NAP}).

\begin{figure}
\centering
\includegraphics[scale=0.5]{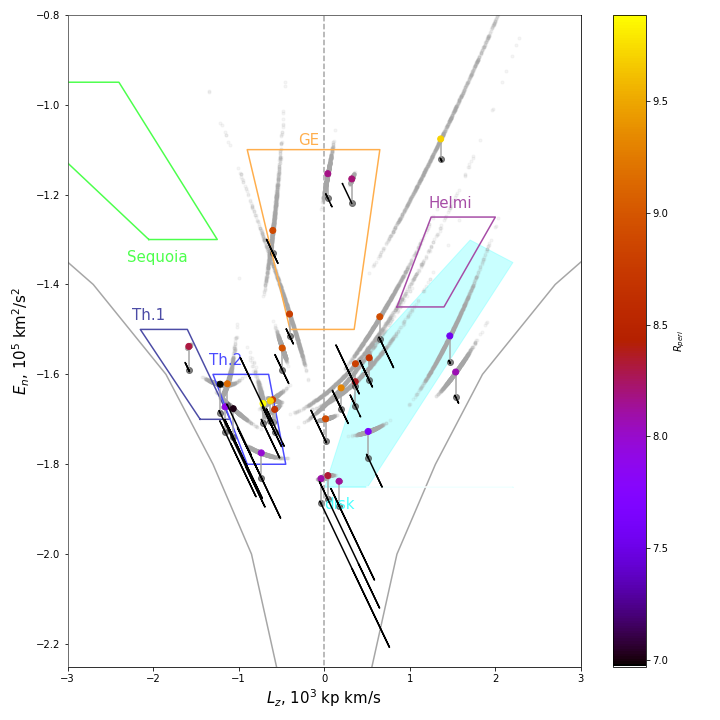}
\caption{Comparison of the position in $E_n - L_z$ diagram calculated in axisymmetric (colored points with blue error bars) and non-axisymmetric (black tilted lines) potentials. $E_n - L_z$ diagram with the loci of the biggest known accretion events: Gaia-Enceladus/Sausage (GE) (\textit{orange}), Helmi stream (\textit{purple}), Sequoia (\textit{green}), Thamnos 1, 2 (\textit{darkblue, blue}); and disk (\textit{cyan}).}
\label{fig:bar}
\end{figure}

In Fig.\ref{fig:bar} we can see that first of all changing the potential modifies the total energy ($E_n$) that stars possess nowadays. Besides, in their motion in the Galactic non-axisymmetric bar potential, both total energy and the angular momentum component along the z-axis evolve with time. Clearly, total energy and angular momentum along the z-axis do not conserve anymore. The Jacobi integral $E_J = E_n - \Omega_bL_z$ is constant instead. As a result of the bulge/bar time evolution $E_n$ anti-correlates with $L_z$ in the same fashion (same slope) for all the stars (\cite{Tkachenko_2023}). For some stars, these changes are large and move them far from the location they had in the axisymmetric potential. This highlights the importance of bar/bulge structure in the dynamical evolution of these stars. However, it is worth mentioning that for the stars we identified to be probable members of  accretion events the change in position due to bar perturbations is not significant. The shifted points in $E_n-L_z$ diagram are still well inside the accretion events' regions (GE, Thamnos 1,2).

For both potentials, we run simulations 1 Gyr long in time with step $\Delta t = 0.1$ Myr. For non-axisymmetric potential, though, we did not calculate uncertainties since it is a highly time-consuming process.

\section{Conclusions}
Three main parameters that can be used to obtain the origin of the star are: kinematics, chemistry and age. In this study, we explored all three of them for a sample of 28 very metal-poor stars for whom we had previously derived ages (\citep{Plotnikova_2022}). Kinematics and chemistry are routinely used to assign stars to different components of the Milky Way galaxy, or to the various accreted systems that have been identified over the years in the Galactic halo.
The basic conclusions of our study can be summarized as follows:

\begin{itemize}
    \item [1.] We have identified a group of stars with clear signatures of Halo population according to their chemistry, age, and kinematics. 
    However, we speculate that several of them, the oldest 8 indeed,  could have formed in the primordial Bulge because of their orbital parameters (large eccentricity and low $Z_{max}$ mostly) and large $\alpha$-abundance.
    \item [2.] We have  identified another group of stars that we tentatively associate with Gaia-Enceladus/Sausage (GE) (\textit{orange}), and Thamnos 1, 2 (\textit{dark blue, blue}) according to their similarity in kinematics, chemistry and age.
\end{itemize}

The identification  of a group, although small, of stars probably belonging to the infant Bulge is rather intriguing. In fact, it supports a scenario in which the first component of the Milky Way to form was the Bulge via a fast collapse {\it a l\`a} ELS (\cite{Eggen_1962}). The other components have then assembled with a major contribution from systems that engulfed into the Milky Way later on, in agreement with the widely accepted merging scheme (\cite{Searle_1978}).

\begin{acknowledgments}
The comments of an anonymous referee have been much appreciated. AP acknowledges the Ulisse program of Padova University which allowed her to spend a period at Concepcion University, where part of this work has been done. Also, AP acknowledges Roman Tkachenko for the useful consultations. SV gratefully acknowledges the support provided by Fondecyt regular n. 1220264 and by the ANID BASAL projects ACE210002 and FB210003. SO acknowledges DOR 2020, University of Padova.
\end{acknowledgments}

\bibliography{Mpoor2}{}
\bibliographystyle{aasjournal}



\end{document}